%% file: main_arxiv.tex
\documentclass[10pt,conference]{IEEEtran}
\IEEEoverridecommandlockouts
\usepackage{cite}
\usepackage{amsmath,amssymb,amsfonts}
\usepackage{algorithmic}
\usepackage{graphicx}
\usepackage{textcomp}
\usepackage{xcolor}
\usepackage{multicol}
\usepackage{multirow}
\usepackage{titlesec}

\usepackage[caption=false,font=footnotesize]{subfig}
\begin{document}

\title{Bridging the Gap Between Simulated and Real Network Data Using Transfer Learning
\thanks{This paper was submitted to IEEE NetSoft 2026.}
}

\author{\IEEEauthorblockN{Carlos Güemes-Palau,
Miquel Ferriol-Galmés,
Jordi Paillisse Vilanova, 
Albert López-Brescó,}
\IEEEauthorblockN{Pere Barlet-Ros and
Albert Cabellos-Aparicio}
\IEEEauthorblockA{
\textit{Barcelona Neural Networking Center, Universitat Politècnica de Catalunya}\\
Barcelona, Spain\\
\{carlos.guemes, miquel.ferriol, jordi.paillisse, albert.lopez, pere.barlet, alberto.cabellos\}@upc.edu }
}
\maketitle

\begin{abstract}
Machine Learning (ML)-based network models provide fast and accurate predictions for complex network behaviors but require substantial training data. Collecting such data from real networks is often costly and limited, especially for critical scenarios like failures. As a result, researchers commonly rely on simulated data, which reduces accuracy when models are deployed in real environments.
We propose a hybrid approach leveraging transfer learning to combine simulated and real-world data. Using RouteNet-Fermi, we show that fine-tuning a pre-trained model with a small real dataset significantly improves performance. Our experiments with OMNeT++ and a custom testbed reduce the Mean Absolute Percentage Error (MAPE) in packet delay prediction by up to 88\%. With just 10 real scenarios, MAPE drops by 37\%, and with 50 scenarios, by 48\%.
\end{abstract}


\setlength{\textfloatsep}{7pt}
\titlespacing*{\section}{0pt}{*1}{*0.95}  
\titlespacing*{\subsection}{0pt}{*1}{*0.95}

\section{Introduction}
\label{sec:introduction}
\input{sections/introduction}

\section{Background}
\label{sec:background}
\input{sections/background_v2}

\section{Problem statement}
\label{sec:problem_statement}
\input{sections/problem_statement}


\section{Methodology and experimental design}
\label{sec:methodology_and_experimental_design}
\input{sections/methodology_and_experimental_design}

\section{Evaluation}
\label{sec:evaluation}
\input{sections/evaluation}

\section{Related work}
\label{sec:related_work}
\input{sections/related_work}

\section{Conclusion}
\label{sec:conclusion}
\input{sections/conclusion}


\section*{Acknowledgments}
This publication is part of the I+D+i project titled BLOSSOMS, grant PID2024-158530OB-I00, funded by MICIU/AEI/10.13039/501100011033/ and by ERDF/EU. This work is also partially funded by the Catalan Institution for Research and Advanced Studies (ICREA).
Carlos Güemes is funded by the AGAUR-FI ajuts (Grant Ref. 2023 F-1 00083) Joan Oró of the Secretariat of Universities and Research of the Department of Research and Universities of the Generalitat of Catalonia and the European Social Plus Fund.

\bibliographystyle{IEEEtran}
\bibliography{main}

\end{document}

%% file: sections/introduction.tex
Network modeling is crucial for reliable communication, traffic optimization, and network design. Traditional approaches, such as Discrete Event Simulation (DES), have been the standard for accurately modeling network behavior.
Tools like NS-3~\cite{Riley2010} and OMNeT++~\cite{Varga2019} simulate every network event. They provide highly detailed reconstructions of network dynamics but at a high computational cost, which limits scalability as networks grow.
Recent advancements in Machine Learning (ML) have introduced transformative possibilities for network modeling, exemplified by models such as MimicNet~\cite{10.1145/3452296.3472926}, DeepQueueNet~\cite{10.1145/3544216.3544248}, and the RouteNet~\cite{ferriolgalmés2022routenetfermi} family.
These innovations address the limitations of DES, enabling faster and more scalable network modeling at the price of a slight reduction in accuracy. 


ML models learn to approximate network behavior accurately and (relatively) cheaply by analyzing traffic patterns in training datasets.
However, this reliance on data introduces significant challenges. To ensure accurate and reliable predictions, training datasets must meet three essential criteria: abundance, diversity, and completeness. Abundance is crucial to expose the model to a wide range of network conditions, enhancing its ability to generalize across scenarios. Diversity refers to the inclusion of a broad spectrum of real-world situations, including rare edge cases that deviate from typical network operations (e.g., highly congested links, heterogeneous traffic patterns, etc). Completeness ensures that all relevant factors influencing network behavior are captured, preventing critical variables from being overlooked.

Nevertheless, acquiring datasets that meet these criteria is commonly challenging. Real-world network data, while accurate, is costly to collect and often incomplete. Publicly available datasets frequently lack the diversity and edge-case scenarios required for robust model training. For instance, CAIDA's Anonymized Internet Traces~\cite{CAIDA} primarily focus on common traffic patterns, limiting their utility for scenarios involving atypical or highly variable network conditions. Testbed networks provide a controlled environment to generate data, but their deployment at scale is prohibitively expensive. These challenges hinder the development of high-quality ML models for real-world network applications.

Researchers often turn to network simulators to generate training datasets to mitigate data scarcity~\cite{10.1145/3452296.3472926, 10.1145/3544216.3544248, ferriolgalmés2022routenetfermi, li2024glancegraphbasedlearnabledigital}. Simulators offer a controlled and flexible environment for producing diverse and rare scenarios without the risks associated with live production networks. Conversely, generating simulated datasets is computationally expensive and often fails to capture the subtle nuances of real-world network dynamics. This arises from idealized assumptions in simulations and the proprietary nature of commercial network hardware. Consequently, ML models trained exclusively on simulated data exhibit reduced accuracy and reliability when deployed in production environments. This observation has been corroborated by other researchers~\cite{10635943} and by our findings in the evaluation (Table~\ref{tab:results}).

This paper studies the viability of using simulated data to train ML models for real-world network behavior prediction. We propose a hybrid approach leveraging transfer learning~\cite{10.5555/2998687.2998769, pmlr-v15-bengio11b} to bridge the gap between simulation and real-world network environments. This consists in training and evaluating our approach by combining simulated network scenarios with real ones (Figure~\ref{fig:transfer_learning_pipeline}). We use RouteNet-Fermi~\cite{ferriolgalmés2022routenetfermi}, a state-of-the-art ML network model, as our reference model in our comparison. However, we believe that our findings are generalizable to other ML-based solutions. We then use our configurable testbed network to record the real network scenarios. In our evaluation, we increase the accuracy of the resulting model by up to 88\% relative to the version that only uses real-world network data. We also find that our approach is extremely data efficient: limiting the training process to 10 recorded network scenarios, using fine-tuning results in a model with $37\%$ less error than when training from scratch.

The rest of the paper is structured as follows: Section~\ref{sec:background} provides an overview of transfer learning, highlighting its relevance to our approach and to RouteNet-Fermi, the ML network model used to evaluate it. Section~\ref{sec:problem_statement} describes the problem statement and introduces the proposed framework, whose details and implementation are then expanded in Section~\ref{sec:methodology_and_experimental_design}. Finally, Section~\ref{sec:evaluation} presents an evaluation of our approach, including key findings and insights.

%% file: sections/background_v2.tex
\subsection{Transfer Learning}

Transfer learning enhances performance in a target task (receiver) by reusing knowledge from a related source task (donor) \cite{10.5555/2998687.2998769, pmlr-v15-bengio11b}. 
In this paper, we specifically employ fine-tuning~\cite{10.5555/2969033.2969197} to transfer useful knowledge from simulated to real-world network scenarios. Fine-tuning is a particularly popular transfer learning technique for neural networks (NNs). It involves reusing a pre-trained donor model by transferring some of its learned weights to initialize the receiver model. These encode knowledge from the donor's training process, providing the new model with a strong starting point, potentially reducing training time and improving accuracy. The benefits are higher if the target dataset is small.



Neural network-based models are typically composed of one or more layers that perform progressively complex transformations of the input data. Lower layers often focus on extracting general features from the input (e.g., basic statistical patterns), while higher layers specialize in learning task-specific features or making predictions~\cite{zeiler2013visualizingunderstandingconvolutionalnetworks}. 
Taking into account this, and how similar donor and receiver tasks are, the receiver model initialize its weights in one of three ways:
\begin{enumerate}
    \item Reused and frozen weights: The weights are transferred from the donor model and remain fixed during training. It decreases the training's computational cost.
    \item Reused and adjustable weights: The weights are transferred from the donor model but are allowed to update during training and to better fit the target.
    \item Randomly initialized weights: The layer's weights are trained from scratch if there is no useful knowledge to transfer from the donor (i.e., task-specific layers).
\end{enumerate}

\subsection{RouteNet-Fermi}

RouteNet-Fermi~\cite{ferriolgalmés2022routenetfermi} is a state-of-the-art network performance model. Specifically, it uses a custom representation of the network along with a modified Message-Passing Neural Network architecture \cite{pmlr-v70-gilmer17a} that exploits the interactions between traffic flows and the underlying devices. As a result, it has proven to generate accurate predictions at a fraction of the computational cost of alternatives such as DES. Its design also makes it robust to infer under unseen topologies, including larger than those seen during the model's training. It was also evaluated with both simulated and real-world network data.


%% file: sections/problem_statement.tex
\begin{figure}[t]
    \centering
    \includegraphics[width=0.85\linewidth]{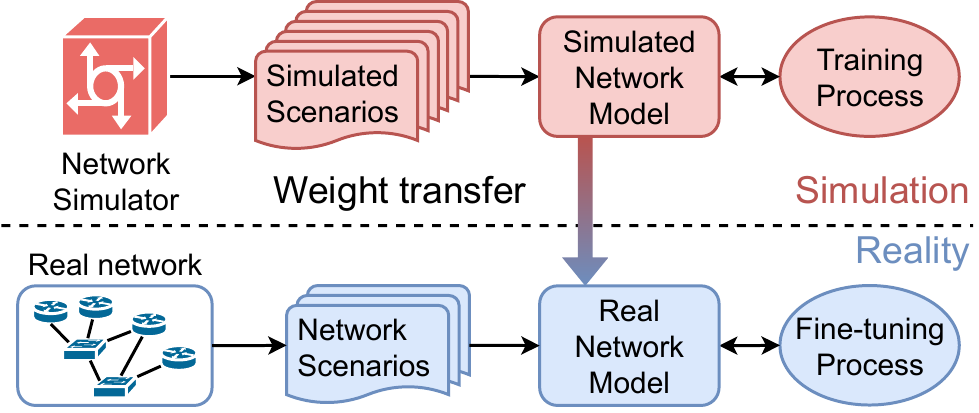}
    \caption{Summary of the proposed hybrid approach. Simulated network scenarios are used to first train a network model. Then, this model is fine-tuned using a smaller dataset from real-world network data.}
    \label{fig:transfer_learning_pipeline}
\end{figure}

Building accurate ML models for network behavior prediction faces a fundamental challenge: the scarcity and diversity of real-world network data. Network data often requires extensive testbed deployments or prolonged production network monitoring. Privacy concerns and hardware limitations further restrict access to critical metrics, such as detailed packet captures or router-specific configurations.
In contrast, simulated data is abundant and diverse but fails to capture the nuances of real-world network dynamics, including hardware-specific behaviors and unexpected edge cases. This mismatch results in models that perform well on simulated scenarios but struggle when applied to real-world networks.

To address this issue, we propose a hybrid approach that combines the strengths of simulated and real-world data using transfer learning, illustrated in Figure~\ref{fig:transfer_learning_pipeline}.
The core idea is to train an ML network model on simulated data and later refine it with real-world network data. We begin by leveraging the diversity and abundance of synthetic scenarios to learn generalized patterns. Using a network simulator, for example, we can generate a comprehensive dataset with diverse network configurations and scenarios. Then, we fine-tune the model by transferring weights from the simulated network model while using a small dataset of real-world network data to adapt it to the specific environment. By doing so, we aim to bridge the gap between simulation and reality, enabling the model to make accurate predictions in real-world scenarios with minimal reliance on extensive real-world datasets.

%% file: sections/methodology_and_experimental_design.tex

Our approach involves two main components: (1) training the RouteNet-Fermi~\cite{ferriolgalmés2022routenetfermi} model on simulated data and (2) fine-tuning it with real-world network data. This methodology enables us to combine the broad generalization capabilities gained from simulation with real-world specificity.

\subsection{Model architecture}

We use a modified RouteNet-Fermi architecture to predict network performance metrics. The architecture can be decomposed into three main blocks:
\begin{enumerate}
    \item Encoding: Multi-layer perceptrons (MLPs) generate initial embeddings for network elements.
    \item Message Passing Algorithm (MPA): The embeddings are refined using the relationships between network elements by employing Gated Recurrent Units (GRU)~\cite{cho2014learningphraserepresentationsusing}.
    \item Readout: The final flow embeddings are used to predict performance metrics via an MLP.
\end{enumerate}

It should be noted that RouteNet-Fermi assumes stationary traffic, a condition that does not always apply to real-world network data. In turn, we adapt RouteNet-Fermi to non-stationary traffic by splitting network scenarios into temporal windows and predicting performance metrics for each window individually. This ensures the stationarity assumption applies only within shorter intervals, allowing the model to adapt to changing traffic conditions as in real-world scenarios.

Adapting the architecture requires two key modifications. First, the input features are adjusted to include window-specific attributes rather than global flow-level parameters. This includes features such as flow bandwidth and packet rate per window. Second, we introduce a GRU neural network during the MPA phase to capture inter-window dependencies. This mechanism updates queue embeddings in each window using those from the previous window, enabling the model to propagate temporal information. Overall, these measures aim to improve accuracy under non-stationary traffic conditions.



\subsection{Manual transfer learning}
\label{sec:proposed_ft}
To summarize, we currently have an ML-model architecture, RouteNet-Fermi, a large dataset of simulated network scenarios, and a small dataset of real-world network scenarios. We propose using transfer learning to leverage the strengths of the simulated dataset while adapting the model to real-world conditions.
We start by training the network model with the simulated data to act as a foundation for the network model for real-network data. When fine-tuning, a critical decision is determining how to handle the weights of each block in the network.
Following the principles outlined in \cite{zeiler2013visualizingunderstandingconvolutionalnetworks}, we evaluate those configurations that adhere to the following guidelines:
\begin{itemize}
    \item Layer dependencies: We avoid configurations where a block is frozen or fine-tuned if preceded by a re-trained block. Otherwise, it would disrupt the natural flow of learned representations. We also avoid configurations where a block is frozen if preceded by a fine-tuned block.
    \item Trainable weights: We never freeze all blocks, as this would leave no trainable parameters for adaptation.
    \item Always transfer something: We never re-train all blocks, as it is equivalent to training the model from scratch.
\end{itemize}

The resulting testbed configurations are listed in Table~\ref{tab:results}. We split the network into blocks rather than individual layers to align with RouteNet-Fermi’s architecture. Unlike sequential NNs like MLPs, RouteNet-Fermi operates more like an ensemble of smaller NNs that work in parallel. For instance, the MLPs in the encoding block process individual network elements independently. Grouping these layers into blocks provides a structured approach to fine-tuning while ensuring that dependencies between blocks are respected. Furthermore, the shallow depth of the RouteNet-Fermi's internal NNs, with the deepest component being a 3-layered MLP, limits the benefit of fine-tuning individual layers.

\begin{figure}[t]
    \centering
    \includegraphics[width=0.8\linewidth]{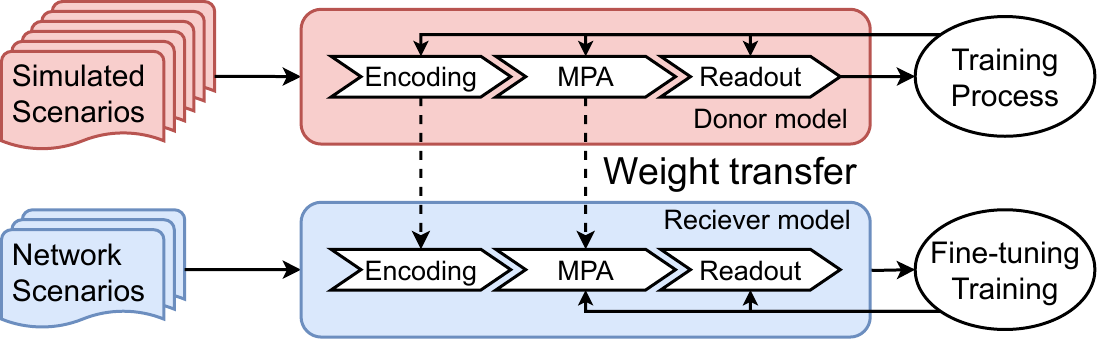}
    \caption{Visual example of fine-tuning a RouteNet-Fermi~\cite{ferriolgalmés2022routenetfermi} model, where the Encoding is frozen, the MPA is fine-tuned, and the Readout is re-trained.}
    \label{fig:fine_tuning_example}
\end{figure}

In Figure~\ref{fig:fine_tuning_example}, we show an example of how the model can be fine-tuned. In this example, the chosen fine-tune configuration was to freeze the Encoding block, fine-tune the MPA block, and re-train the Readout block. As a result, we only transfer the Encoding and MPA weights from the donor model, while the Readout block's weights are randomly initialized as in traditional training. Then, during the fine-tuning training, the Encoding block is excluded so as not to modify its weights. Note that the fine-tuning training is otherwise similar to the original training process, but using the real-network samples and with a diminished learning rate ($\approx10\times$  smaller).

\subsection{Automated transfer learning}
\label{sec:automated_ft}

In addition to the previous manual configurations, we also test our approach using automated fine-tuning approaches from the state-of-the-art. These do not require manually deciding which blocks to freeze, fine-tune, or retrain, reducing trial-and-error:

\begin{itemize}
    \item Autofreeze~\cite{liu2021autofreeze}: This method begins by loading all donor weights and setting them as trainable. During training, blocks whose weight gradients fall under a threshold are frozen, minimizing computational costs.
    \item L2-SP~\cite{pmlr-v80-li18a}: This method consists of adding a regularization term in the loss function involving the L2-distance between the receiver and donor weights. This guides the learning of the receiver model and is more effective at avoiding overfitting than the standard L2 regularization.
    \item GTOT-Tuning~\cite{zhang2022fine}: A more advanced version of L2-SP meant for Graph Neural Networks (GNNs). Instead of comparing weights, it compares node embeddings after the MPA using the Masked Wasserstein Distance, which incorporates relational information.
\end{itemize}

\subsection{Testbed}
\label{subsec:testbed}
\begin{figure}[t]
    \centering
    \includegraphics[width=0.90\linewidth]{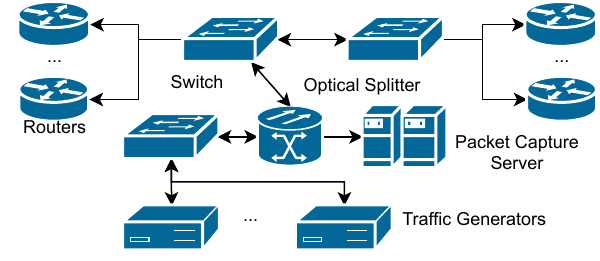}
    \caption{Diagram summarizing the testbed's structure.}
    \label{fig:testbed}
\end{figure}

To collect real network samples, we use a custom testbed with up to 8 real routers and traffic generators connected via switches. VLAN-based configurations allow emulation of diverse network topologies. Links range from 1 to 40 Gbps to simulate modern network conditions. Traffic is generated using 2–8 servers running MGEN and Tcpreplay. An optical splitter enables passive traffic capture for analysis, ensuring low interference. Figure~\ref{fig:testbed} shows the testbed's structure.


%% file: sections/evaluation.tex
In this section, we evaluate the effectiveness of our proposed fine-tuning approach. The primary goals of this evaluation are to answer the following key questions:
\begin{enumerate}
    \item Can fine-tuning be used to create a more accurate model than one solely trained with scarce real-world network data? If so, which configuration of weight handling (freezing, fine-tuning, or re-training) works best?
    \item How does the availability of real-network data influence the impact of transfer learning?
\end{enumerate}

To address these questions, we evaluate the model’s performance improvements when transitioning from a model trained solely on real-world network data to one that uses both types of data. The evaluation is performed across various fine-tuning configurations and under different network conditions.
The simulated dataset is generated using the OMNeT++\cite{Varga2019} simulator and includes network scenarios with topologies ranging from 5 to 8 nodes and diverse routing configurations. It provides a broad and varied training foundation, covering a wide range of network conditions to enable the model to learn and adapt to diverse network configurations. The real-world network dataset is extracted from the testbed described in Section \ref{subsec:testbed}, configured as a fixed 5-router topology. The Poisson and On/Off distributions follow a static routing configuration, while the MAWI distribution considers multiple routing configurations to increase the number of available samples (needed to evaluate question 2).

Both datasets consist of network scenarios, each comprising between 177 thousand and 3.6 million packets successfully sent. When predicting their behavior, they are aggregated into fixed-sized temporal windows, each 100ms long. This windowing approach can capture non-stationary traffic patterns and facilitate meaningful comparisons between simulated and real-world conditions. Furthermore, both datasets contain traffic flows based on three distinct distributions: Poisson, On/Off, and MAWI. Poisson traffic represents steady patterns characterized by exponentially distributed inter-arrival times. On/Off traffic alternates between idle periods and constant bit rate transmissions, capturing bursty network behavior. The MAWI distribution, on the other hand, follows the packet inter-arrival distribution measured in the internet traces published at the MAWI Working Group Traffic Archive~\cite{10.5555/1267724.1267775}.

\begin{table}[t]
\centering
\caption{Number of network scenarios available.}
\resizebox{\columnwidth}{!}{%
\begin{tabular}{@{}lcccccc@{}}
\hline
\multirow{2}{*}{Partition} & \multicolumn{3}{c}{Simulated Scenarios} & \multicolumn{3}{c}{Real Scenarios} \\
            & Poisson        & On/Off        & MAWI       & Poisson       & On/Off        & MAWI       \\ \hline
Training    & 3145           & 3121          & 1634         & 30            & 38            & 165           \\
Validation  & 628            & 624            & 326          & 6             & 7             & 33            \\
Evaluation  & -              & -              & -            & 4             & 5             & 22            \\ \hline
\end{tabular}%
}
%
\label{tab:ds_size}
\end{table}

Table~\ref{tab:ds_size} presents the number of network scenarios available for our study, categorized by source (simulated or real), traffic distribution (Poisson, On/Off, or MAWI), and partition (training, validation, or evaluation).
Notably, the simulated dataset is significantly larger than the real-world dataset, reflecting the limited availability in real-world network data. 

As part of the evaluation, we establish two baseline models: one trained exclusively on simulated scenarios and the other solely on real-world scenarios. We then use the simulated data only model for both manual and automated fine-tuning. Manual fine-tuning configurations and criteria for selecting these configurations are detailed in subsections~\ref{sec:proposed_ft}. The automated fine-tuning approaches are listed in Section~\ref {sec:automated_ft} and have been implemented according to their original papers, including hyperparameter values, with necessary adaptations made to fit RouteNet-F's architecture as needed.
Our windowed implementation of the RouteNet-Fermi model is built upon their public repository\footnote{https://github.com/BNN-UPC/RouteNet-Fermi}, using TensorFlow 2.15. Model hyperparameters (e.g., embedding sizes) are borrowed directly from the original implementation. All models ran for a redundant number of epochs until the validation error stopped improving, and kept the weights that minimized it. Additional implementation details are available in our public repository\footnote{https://github.com/BNN-UPC/Papers/wiki/Bridging-the-Gap-Between-Simulated-and-Real-Network-Data-Using-Transfer-Learning}.

\subsection{Fine-tuning the network model}

We evaluate the impact of fine-tuning by building a network model to predict the average packet delay in each temporal window for every flow. The predictions are assessed using the Mean Absolute Percentage Error (MAPE), a relative error metric. The evaluation involves first training a baseline model solely on simulated data to serve as the donor model, and then fine-tuning the pre-trained model using real-world network data. This is done both in the Poisson and On/Off scenarios.

\begin{table}[t]
\centering
\caption{Results of all fine-tuning configurations. Models are tasked to predict the mean packet delay for each flow-window pair. Models are evaluated using the normalized MAPE, relative to the model without fine-tuning. Values under 1 indicate improvement; lower is better, and highlighted result is best.}
\resizebox{\columnwidth}{!}{%
\begin{tabular}{ccccc}
\hline
\multicolumn{3}{c}{Fine-tuning configuration} & \multicolumn{2}{c}{Normalized Delay MAPE}               \\
Encoding      & MPA           & Readout       & Poisson        & On/Off \\ \hline
Freeze        & Freeze        & Fine-tune     & 0.557          & 0.997                      \\
Freeze        & Freeze        & Re-train      & 1.056          & 1.333                      \\
Freeze        & Fine-tune     & Fine-tune     & 0.273          & 0.448                      \\
Freeze        & Fine-tune     & Re-train      & 0.180 & 0.406             \\
Freeze        & Re-train      & Re-train      & 0.180 & 0.406             \\
Fine-tune     & Fine-tune     & Fine-tune     & 0.283          & 0.434                      \\
Fine-tune     & Fine-tune     & Re-train      & 0.474          & 0.614                      \\
Fine-tune     & Re-train      & Re-train      & 0.474          & 0.614                      \\ \hline
\multicolumn{3}{c}{Autofreeze~\cite{liu2021autofreeze}}       & 0.131         & \textbf{0.197} \\
\multicolumn{3}{c}{L2-SP~\cite{pmlr-v80-li18a}}       & 0.312          & 0.434                     \\ 
\multicolumn{3}{c}{GTOT-Tuning~\cite{zhang2022fine}}       & \textbf{0.119}          & 0.207  \\ \hline
\multicolumn{3}{c}{Simulated data only}       & 9.900          & 17.958                     \\ \hline
\end{tabular}%
}
\label{tab:results}
\end{table}

Table~\ref{tab:results} presents the MAPE for all evaluated fine-tuning configurations. Models are tasked to predict the average packet delay in each flow-window pair. The manual configurations involve varying the treatment of weights (to freeze, fine-tune, or re-train) in the three main blocks of the RouteNet-Fermi model: Encoding, MPA, and Readout. We also show the performance of automated finetuning methods and training exclusively on simulated data. The results are normalized against a model trained with the real-world network data only, without fine-tuning. That is, a value of 1 indicates the same accuracy as the baseline, while lower values indicate an improvement over the baseline. 

Addressing the first question raised at the beginning of the evaluation, the results demonstrate that \textit{fine-tuning consistently improves the accuracy of the network model}. All but one of the manual configurations tested reduce the model’s error across both traffic distributions compared to the baselines. The most effective configuration consists of \textit{freezing the Encoding block and re-training the Readout block}. It achieves an $82\%$ improvement over the real-world network baseline for Poisson traffic and a $59.4\%$ improvement for On/Off traffic. In this case, \textit{both fine-tuning and re-training the MPA block return similar accuracy}, showing that training correctly adjusts its weights independently of whether there is a transfer or not. 

\begin{figure}[t] 
    \centering
  \subfloat[Poisson\label{fig:poisson_pdf}]{%
       \includegraphics[width=\linewidth]{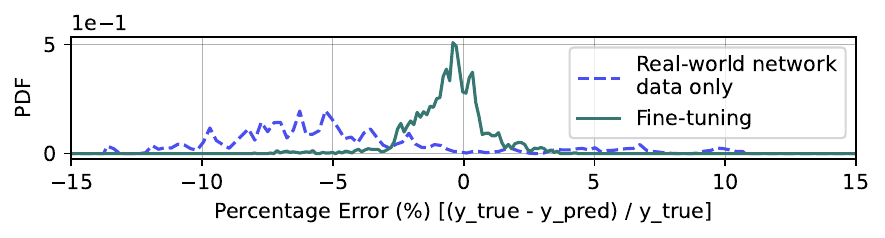}}
  \\ \vspace{-4mm}
  \subfloat[On/Off\label{fig:on_off_pdf}]{%
        \includegraphics[width=\linewidth]{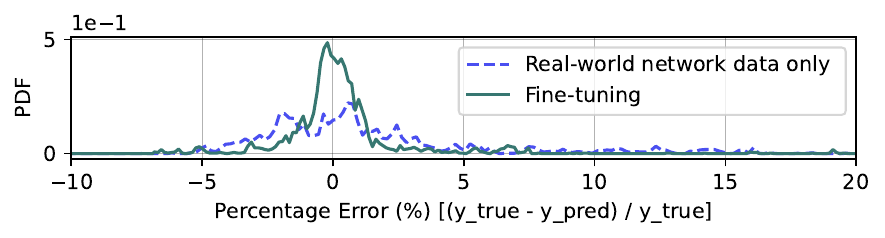}}
  \caption{Probability Density Function (PDF) of percentage error when predicting the mean packet delay of each flow-window pair. It compares the best manual fine-tuning configuration (freezing Encoding, fine-tuning MPA, re-training Readout) against the real-world network data only model.}
  \label{fig:pdfs} 
\end{figure}

To further illustrate the differences, Figure~\ref{fig:pdfs} illustrates Probability Density Functions (PDFs) representing the relative error distributions obtained by the best fine-tuning configuration and the real-world network data only model. 
In them, it is clear that the fine-tuned model's predictions are generally closer to having a 0\% error and exhibit a balance between underestimation and overestimation. In contrast, the real-world network data only model shows a clear bias in the Poisson dataset, favoring overestimation, with most predictions displaying an average relative error of approximately $-5\%$. 
However, one manual configuration results in a performance decline across both distributions: freezing the Encoding and MPA blocks while re-training the Readout block. This underscores the potential for \textit{negative transfer}, where knowledge from the donor model hinders rather than enhances the receiver model's performance when not applied correctly.

The automated fine-tuning solutions also proved effective, with both \textit{Autofreeze~\cite{liu2021autofreeze} and GTOT-Tuning~\cite{zhang2022fine} surpassing the best manual configuration}.
Specifically, the former improves MAPE by $80\%$ over the baseline for On/Off traffic and the latter $88\%$ for Poisson traffic, cementing the value of using simulated data and fine-tuning to enhance model accuracy

Furthermore, models trained exclusively on simulated data showed significantly higher error rates: $9.90$ and $17.96$ times larger than the real-world network data only model in Poisson and On/Off traffic, respectively. This reaffirms that simulated data alone is insufficient for accurate real-world predictions, aligning with previous research~\cite{10635943}. Nonetheless, the success of fine-tuning demonstrates that simulated data can still provide valuable insights. Otherwise, our approach would not outperform models trained only on real-world network data.

\subsection{Impact of real-world network data availability}

\begin{figure}[t]
    \centering
    \includegraphics[width=\linewidth]{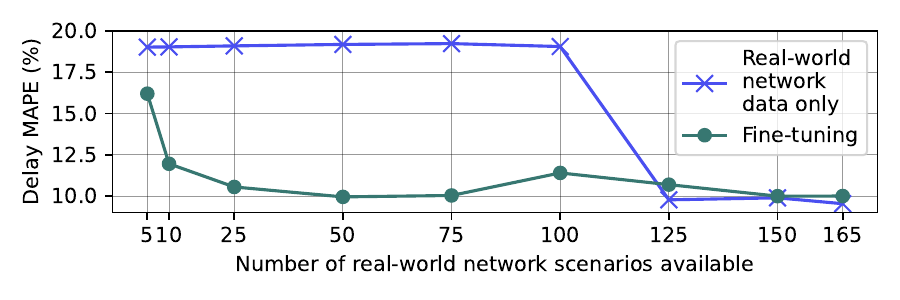}

    \caption{Impact of real-world network data availability on the resulting model's MAPE when predicting the average packet delay in each flow-window pair. Network scenarios belong to the MAWI distribution. }
    \label{fig:fine_tune_size}
\end{figure}

In this section, we evaluate how the availability of network data impacts the benefits of transfer learning compared to training models from scratch. For this experiment, we use internet traffic traces from the MAWI distribution. Figure~\ref{fig:fine_tune_size} presents the model’s MAPE when predicting the average delay per flow-window pair with varying amounts of real-world network data. The fine-tuned model follows the best-performing manual configuration from Table~\ref{tab:results}: freezing the Encoding block, fine-tuning MPA, and re-training Readout. 

The results show that with only five scenarios available, fine-tuning reduces the model’s MAPE from $19.01\%$ to $16.19\%$. \textit{Increasing the number of available scenarios to 10 further decreases the MAPE to $11.95\%$, a $37\%$ reduction. With 50 scenarios, the fine-tuned model achieves its largest improvement, lowering the MAPE to $9.95\%$, a $48\%$ decrease.} This demonstrates the data efficiency enabled by fine-tuning.
However, these results also reveal their limitations: beyond 125 scenarios, fine-tuning provides no additional benefit over training with real-world network data alone. While this threshold may vary between models, RouteNet-Fermi has demonstrated the ability to achieve high accuracy even with limited training data~\cite{ferriolgalmés2022routenetfermi}. This reinforces the idea that fine-tuning is particularly valuable when real-world network data is scarce.

%% file: sections/related_work.tex
\subsection{Network modeling}
State-of-the-art network models can be broadly categorized into two fields: DES and ML-based models. DES simulators such as OMNeT++\cite{Varga2019} and NS3\cite{Riley2010} have dominated the field, providing highly detailed simulations of network behavior. 
In contrast, ML-based models offer a scalable alternative to DES by leveraging data-driven approaches. GNNs have emerged as the dominant architecture in this space \cite{ wang2022xnet, li2024glancegraphbasedlearnabledigital}, with the RouteNet family of models \cite{ferriolgalmés2022routenetfermi} leading the way. DES-ML hybrids aim to accelerate DES by replacing specific simulation components with ML-based approximations. Notable examples include MimicNet~\cite{10.1145/3452296.3472926}, DeepQueueNet~\cite{10.1145/3544216.3544248} and m3~\cite{10.1145/3651890.3672243}.
However, these models are typically trained on simulated data, limiting their accuracy and reliability in real-world networks.

\subsection{Transfer learning in network modeling}
The application of transfer learning in network modeling remains relatively limited.
For instance, in~\cite{10635943}, the authors employ a Neural Processes (NPs) architecture trained through a transfer learning-based network model using simulated and real network data. While effective, NPs are inherently limited in capturing the relational information present in network scenarios, unlike GNNs such as RouteNet.
In GLANCE~\cite{li2024glancegraphbasedlearnabledigital}, authors employ fine-tuning, but for transferring knowledge between performance metrics rather than addressing the challenges of adapting models to real-world network scenarios.
Transfer learning is more prevalent in fields adjacent to network modeling, such as traffic prediction~\cite{8667446}, intrusion detection~\cite{MAHDAVI2022109542}, and energy consumption reduction~\cite{8879693}.


%% file: sections/conclusion.tex
In this paper, we studied how transfer learning can bridge the gap between simulated and real-world networks. By leveraging simulated data to complement scarce real-world network data, we explored the potential of transfer learning to build enhanced network models. Across our evaluation, we have found that using this approach has resulted in an $88\%$ and $80\%$ error reduction in flows following a Poisson and On/Off distribution, respectively. When instead replicating captured internet traffic, our approach resulted in a $48\%$ error reduction using only 50 recorded network scenarios. When only having 10 scenarios available, the error was still reduced by $37\%$. Furthermore, these benefits apply to both automated transfer learning methods and manual fine-tuning configurations. In conclusion, our approach reduces reliance on extensive real-world datasets, making it both practical and efficient, while benefiting from the strengths of network simulation.
